\documentclass[11pt,twoside]{article}
\usepackage{asp2010}

\resetcounters

\renewcommand{\volume}{TBD}
\renewcommand{\cpryear}{2012}
\renewcommand{\voltitle}{ADASS XXI}
\renewcommand{\volauthor}{Ballester, P. and Egret, D.}

\begin{document}

\title{Benchmarking the CRBLASTER  Computational Framework on a 350-MHz 49-core Maestro Development Board}
\author{Kenneth J.~Mighell$^1$}
\affil{$^1$National Optical Astronomy Observatory, 950 N. Cherry Ave., Tucson, AZ 85719~~U.S.A}

\begin{abstract}
I describe the performance of the CRBLASTER computational framework on a 350-MHz 49-core Maestro Development Board (MDB). The 49-core Interim Test Chip (ITC) was developed by the U.S. Government and is based on the intellectual property of the 64-core TILE64 processor of the Tilera Corporation. The Maestro processor is intended for use in the high radiation environments found in space; the ITC was fabricated using IBM 90-nm CMOS 9SF technology and Radiation-Hardening-by-Design (RHDB) rules. CRBLASTER is a parallel-processing cosmic-ray rejection application based on a simple computational framework that uses the high-performance computing industry standard Message Passing Interface (MPI) library. CRBLASTER was designed to be used by research scientists to easily port image-analysis programs based on embarrassingly-parallel algorithms to a parallel-processing environment such as a multi-node Beowulf cluster or multi-core processors using MPI. I describe my experience of porting CRBLASTER to the 64-core TILE64 processor, the Maestro simulator, and finally the 49-core Maestro  processor itself. Performance comparisons using the ITC are presented between emulating all floating-point operations in software and doing all floating point operations with hardware assist from an IEEE-754 compliant Aurora FPU (floating point unit) that is attached to each of the 49 cores. Benchmarking of the CRBLASTER computational framework using the memory-intensive L.A.COSMIC cosmic ray rejection algorithm and a computational-intensive Poisson noise generator reveal subtleties of the Maestro hardware design. Lastly, I describe the importance of using real scientific applications during the testing phase of next-generation computer hardware; complex real-world scientific applications can stress hardware in novel ways that may not necessarily be revealed while executing simple applications or unit tests.
\end{abstract}

\section{Introduction}

The 49-core Maestro processor Interim Test Chip (ITC) was developed by the U.S. Government and is based on 
the intellectual property of the 64-core TILE64 processor of the Tilera Corporation\footnote{Tilera Corporaton website: http://www.tilera.com}
The Maestro processor is intended for use in the high radiation environments found in space; 
the ITC was fabricated using IBM 90-nm CMOS 9SF technology and Radiation-Hardening-by-Design (RHDB) rules.
The Maestro processor has 49 tiles (cores) arranged in a 7 by 7 mesh on a chip (see Fig.~1).
Each Maestro tile has hardware assist
from an IEEE-754 complaint Aurora FPU (floating point unit) which can achieve speedups of an order of magnitude or
better in floating-point intensive computations as compared to when all floating point operations are done in software.

\begin{figure}[t]
\centering
\includegraphics[scale=0.43]{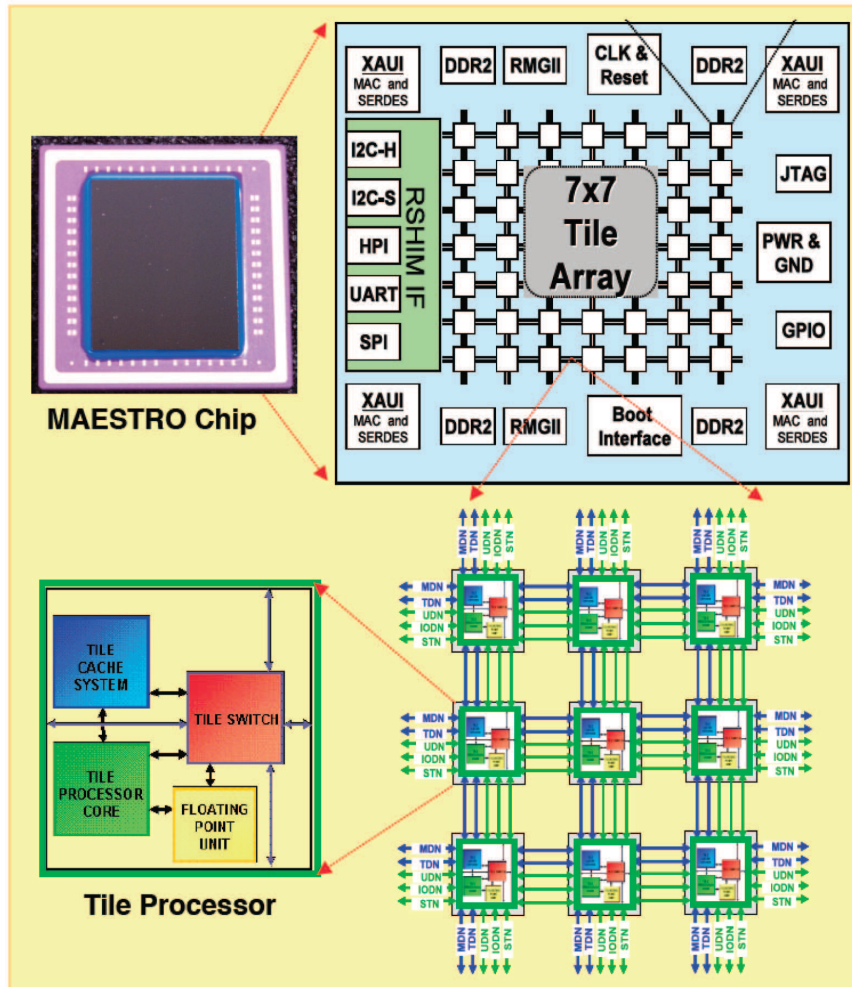}
\caption{Schematic diagram of the 49-core Maestro processor. Credit: Boeing SSED (Solid-State Electronics
Development).}
\end{figure}

CRBLASTER\footnote{CRBLASTER website: http://www.noao.edu/staff/mighell/crblaster/ }
is a parallel-processing computational framework that has been designed to be used by research
scientists for the easy development of image-analysis programs based on embarrassingly
parallel algorithms \citep{Mighell2010}.
The framework is written in the C computer language and parallel-processing is achieved using the high-performance computing industry standard Message
Passing Interface (MPI).  Applications based on the CRBLASTER framework can run
on on multiple cores within a multicore processor and/or
multiple processors within a Beowulf cluster environment.

The first embarrassingly parallel image-analysis algorithm to be implemented with\-in the CRBLASTER
computational framework was 
the L.A.COSMIC algorithm for cosmic-ray rejection in CCD (charge-coupled devices) images
using Laplacian edge detection \citep{vanDokkum}.
Mighell developed a parallel-processing implementation of that algorithm
which splits the computational work across multiple processors that repair unique subsections of the input image.
With all 8 cores in the dual quad-core 2.8 GHz Intel Xeon processors on an Apple Mac Pro computer,
CRBLASTER takes 0.875 seconds 
to process the standard 800$\,\times\,$800 WFPC2 test image
--- a speedup of 50.3 times over van Dokkum's 
{\tt{lacos\_im.cl}} IRAF \citep{Tody1986,Tody1993} script.

The CRBLASTER framework uses two image partitioning algorithms. 
The one dimensional (1D) algorithm partitions an image into $N$ horizontal or vertical image slices.
The two dimensional (2D) algorithm splits an image into $N$ quadrilateral subimages. 
With both partitioning algorithms, the subimages contain approximately $1/N$th of the input image.
Each subimage contains an additional overlap region of pixels beyond all joint partition edges;
in the case the the L.A.COSMIC algorithm, the overlap region is 6 pixels deep.
Computational efficiency is improved by having the fewest overlap pixels possible;
this happens when subimages are nearly square for overlap depths greater than zero.

The master (director) node (core/tile) reads FITS 
(Wells et al.~\citeyear{Wells+1981}, IAU FITS Working Group\footnote{
International Astronomical Union Fits Working Group website:
http://fits.gsfc.nasa.gov/iaufwg/})
images from disk using the CFITSIO\footnote{CFITSIO website:
http://heasarc.nasa.gov/docs/software/fitsio/fitsio.html} library.
The master node splits the input image into $N$ subimges and
then sends them to the worker (actor) nodes.
Once a worker node has finished processing their subimage
(e.g., removing cosmic-ray damage using the L.A.COSMIC algorithm),
the output  subimage is sent to the master node.
The master node gathers up all of the processed subimages.
When all the processed  subimages have been received, the master node
combines them together to make an output image that is the same size 
as the original image.
The combined output image is then written to disk as a FITS file.

\section{Porting CRBLASTER to the TILE64 Platform}

The Tilera 700-MHz TILE64 processor on the TILExpress-20G card 
features 64 identical processor cores (tiles) interconnected
in an 8$\,\times\,$8 mesh architecture.  The TILE64 is programmable
in the ANSI C and C++ languages 
and runs the SMP (Symmetric Multi-Processors) 
Linux operating system. Every tile can run its own instance
of a full operating system.  Any group of tiles can be combined together
using a multiprocessing operating system like SMP Linux. 
The TILE64 processor is energy-efficient; total energy consumption 
with all cores running full application is typically 15 to 22 W.
It is important to note that floating-point operations are done in software
as the TILE64 processor has no hardware assist for floating-point
operations. 

The CRBLASTER framework was ported to the Tilera
64-core TILE64 platform in just 8 hours spread over a few
days using a Tilera TILExpress-20G PCIe card.

The biggest problem encountered with the port of CBLASTER to the
TILE64 platform was the fact the this was a new computer architecture
for the CFITSIO library. The library needs to know the size of C 
{\tt\small{long}} variables
and whether byte swapping is required (endianness).  On its first pass,
CFITSIO guessed the wrong values for these important compiler options.
Three preprocessor macros in the include file
{\tt{fitsio2.h}} were redefined as follows:\\[\parskip]

\noindent
\begin{minipage}[h]{\linewidth}
\footnotesize
\begin{verbatim}
/* MIGHELL 2009SEP23: Tilera Tile64 processor values */
#ifdef __tile__
#undef  MACHINE
#define MACHINE  OTHERTYPE
#undef  BYTESWAPPED
#define BYTESWAPPED TRUE
#undef  LONGSIZE 
#define LONGSIZE 32
#endif

\end{verbatim}
\end{minipage}
With these new macro values, the CFITSIO library builds correctly.
With a good build for the CFITSIO library, 
CRBLASTER was successfully built without encountering any further
problems.

The performance of the CRBLASTER framework 
with the TILE64 processor has
been previously described \citep{Mighell2010}.

\section{Porting CRBLASTER to the TILE64 Simulator}

An important feature of Tilera's Multicore Development Environment (MDE) tool
suite is the availability of a complete system simulator for the TILE64 platform.
The simulator emulates the operations of the TILE64 processor in 
{\em{software}}
not hardware.  Needless to say, this can be very slow.

A 50$\,\times\,$50 subimage was extracted from the standard 800$\,\times\,$800
CRBLASTER test image; this subimage was 256 times smaller than the standard
test image.  Using the smaller test image, the TILE64 simulator took 945 seconds using a single tile.
The total slowdown factor between the TILE64 simulator and the TILE64 processor was a factor of 157.5; 
it was estimated
that using the standard test image would have taken 3.3 days on the TILE64 simulator.

\section{Porting CRBLASTER to the Maestro Simulator}

The Maestro simulator is based on the TILE64 simulator.  
CRBLASTER was ported to the Maestro Simulator in just 6 hours.
Using the 50$\,\times\,$50 test image, the Maestro Simulator took 876 seconds.
The results were verified to be correct; 
the output file was an exact copy of the ``gold standard'' output file.

The Maestro Cross Compiler is much closer to the ANSI/ISO C standard than 
the Tilera Cross Compiler which has many useful C99 features.  Once the 25 lines of code with C99 features
were modified to become standard C90 code, CRBLASTER compiled without any further problems.

Using the standard 800$\,\times\,$800 test image, the Maestro Simulator took 9.09 hours using a single tile.
The results were verified to be correct; 
the output file was an exact copy of the ``gold standard'' output file.

The total time spent porting CRBLASTER to the Maestro Simulator was only 24 hours
spread over a few days.  This was a remarkably quick time considering the novel nature
of the TILE64/Maestro computer architecture and hardware.

\section{Porting CRBLASTER to an Early 100-MHz Maestro Development Board}

CRBLASTER was ported to an early 100-MHz Maestro Development Board (MDB) in February 2011.
The MDB was located the Arlington, Virginia site of the University of Southern California's
Information Sciences Institute (USC/ISI).

The Maestro processor has a 10/100/1000 Ethernet port, but unfortunately it was undergoing testing
and was not available during the port process.  With normal networking to/from the Maestro
processor being disabled, communication using a secure shell client like ssh was not possible.
Bootroms were uploaded using a UART (Universal Asynchronous Receiver/Transmitter) with
an upload speed of 115200 BAUD (14,400 bytes per second); load times were long
-- typically 20 minutes. Access to the machine is 
provided through a text console connected to the UART.

Although each core (tile) on the Maestro processor has an integrated IEEE-754 compliant Aurora FPU,
the FPUs were undergoing testing during the port process.  So at that time, all applications which did floating point
operations needed to have those operations emulated in software.  TILE64 code executes on a Maestro
processor --- just more slowly than Maestro code because of the need to emulate FPU operations.

CRBLASTER running on the 100-MHz Maestro ITC
--- with floating point operations emulated in software --- typically achieved speedup factors
of 20 or more when running on 30 or more tiles.
The total time spent porting CRBLASTER to the Maestro processor was about 20 hours spread over
several days.  
More details of the port to the Maestro processor
can be found in Mighell (\citeyear{Mighell2011}).

\section{Results using a 350-MHz Maestro Development Board}

The 100-MHz MDB was replaced with a 350-MHz MDB with working FPUs for all 49 tiles (see Fig.~2).
Since Ethernet connectivity was still being tested, 
uploading of the bootrom using the UART was still required. 

\begin{figure}[tb]
\centering
\includegraphics[scale=0.39]{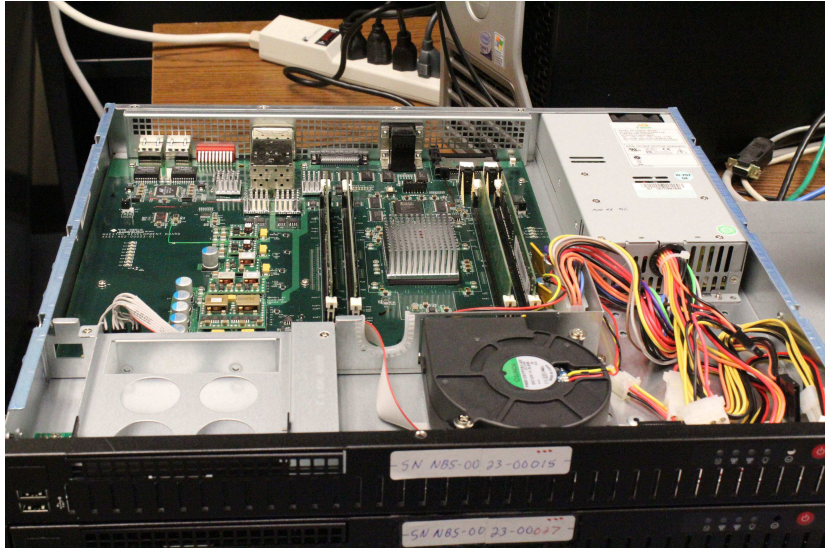}
\caption{The 350-MHz Maestro Development Board (MDB) at the Arlington, Virginia site of USC/ISI.
The cover was removed to expose the internal components.
The 49-core Maestro processor is underneath the heatsink with the thin vertical fins
seen in the center of the image.
Photo courtesy of Jinwoo Suh.}
\end{figure}

Two errors were found in the Maestro Cross Compiler: one was purely software and the
other was related to the FPU.
Working with the team at the Arlington, Virginia site of USC/ISI, functional workarounds were developed.  
Analysis of the differences between the assembly code
produced by the buggy and the workaround versions
lead to improvement of the cross compiler.

The standard 750$\,\times\,$750 pixel test image \citep{Mighell2010}
was used to measure the performance of CRBLASTER with 1 to 45 tiles.
Figure 3 shows the speedup factor of CRBLASTER as a function of the number of processors 
($N$) on a log-log plot and the inset graph shows the
computational efficiency  as a function of $N$.
The speedup factor with $N$ processors is defined as $S_N \equiv t_1/t_N$,
where $t_1$ and $t_N$ are the execution times for 1 and $N$ processors, respectively.
The computational efficiency for $N$ processors is defined as $\epsilon \equiv S_N/N$.
The open diamonds show the performance of the CRBLASTER computational framework
using the L.A.COSMIC algorithm and the 2-D image partitioning method.
The speedup of CRBLASTER using the L.A.COSMIC algorithm with 
9, 24, and 45 tiles was 
5.3, 11.2, and 13.9, respectively.
The computational efficiency with 9, 24, and 45 tiles was, respectively,
59.2\%, 46.8\%, and  30.9\%.

\begin{figure}[t]
\centering
\includegraphics[scale=0.51]{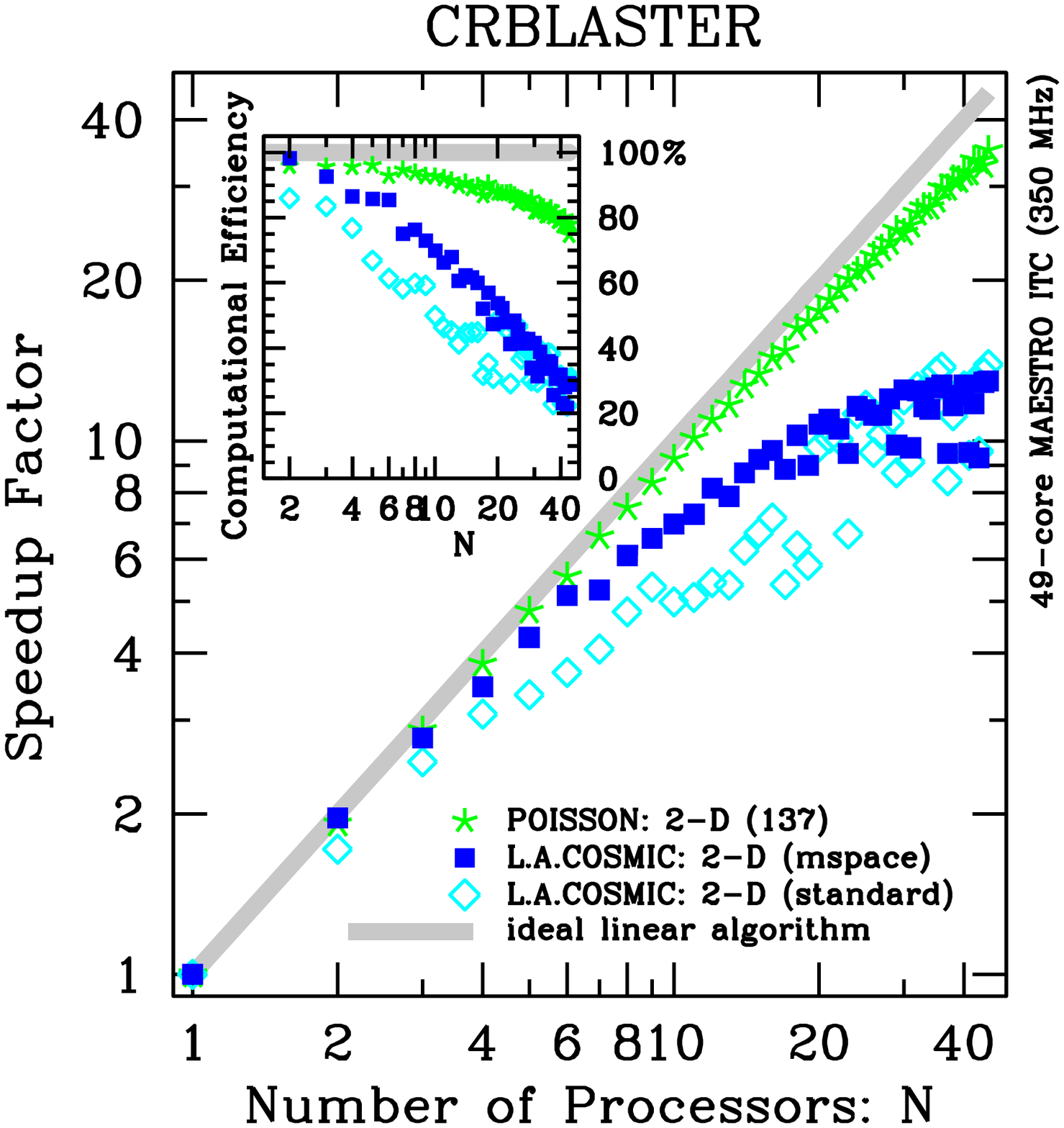}
\caption{Measured performance of CRBLASTER using 1 to 45 cores (tiles) with a 350-MHz 49-core 
Maestro Development Board (MDB) with a Maestro Interim Test Chip (ITC). 
See the text for more details.
}
\end{figure}

In an attempt to boost computational efficiency, Jinwoo Suh (USC/ISI) and I investigated
the use of user-defined heap memory management.
The C library libc from Tilera (\citeyear{Tilera2010}) provides user-defined memory heaps
that can be allocated to any of the four memory controllers on the Maestro processor.
The standard Tilera malloc.h include file has {\em{mspace}} functions that are based on 
Doug Lea's dlmalloc.c\footnote{dlmalloc website: ftp://g.oswego.edu/pub/misc/malloc.c}
user-defined heap memory management code.
The standard CRBLASTER code was modified so 
the operating system and the Hypervisor had exclusive use of memory controller 0
and the master (director) and worker (actor) tiles
shared memory controllers 1, 2, and 3 based on their MPI rank value.
The standard memory allocation function calls, {\tt\footnotesize{malloc()}} and {\tt\footnotesize{calloc()}}, 
and the deallocation function, {\tt\footnotesize{free()}}, were replaced
with their {\em{mspace}} equivalents {\tt\footnotesize{mspace\_malloc()}}, {\tt\footnotesize{mspace\_calloc()}}, 
and {\tt\footnotesize{mspace\_free()}}, respectively.
Replacement was done automatically using the C preprocessor and the following include file:\\[1em]
\begin{minipage}[h]{\linewidth}
\footnotesize
\begin{verbatim}
/* file://msp.h */
#ifndef MSP_H
#define MSPACE_USE
#ifdef MSPACE_USE
#include <malloc.h>
/* msp must be global! */
#ifdef IS_MAIN
mspace *msp = NULL; 
#else
extern mspace *msp; 
#endif /* IS_MAIN */
#define malloc(x) mspace_malloc(msp,(x))
#define calloc(x,y) mspace_calloc(msp,(x),(y))
#define free(x) mspace_free(msp,(x))
#endif /* MSPACE_USE */
#define MSP_H
#endif /* MSP_H */
/* EOF */
\end{verbatim}
\end{minipage}
Two statements were inserted after the last include file statement in the main function:\\[1em]
\begin{minipage}[h]{\linewidth}
\footnotesize
\begin{verbatim}
#define IS_MAIN
#include "msp.h"

\end{verbatim}
\end{minipage}
The following single statement was then inserted after the last include file statement
in the remainder of the .c source files:\\[1em]
\begin{minipage}[h]{\linewidth}
\footnotesize
\begin{verbatim}

#include "msp.h"

\end{verbatim}
\end{minipage}
The following {\em{mspace}} infrastructure code was then inserted in the main function after
the MPI initialization code:\\[1em]
\begin{minipage}[h]{\linewidth}
\footnotesize
\begin{verbatim}

{ /* mspace infrastructure */
    int loc; /* memory controller to be used: 0, 1, 2, 3 */
    alloc_attr_t attr;
    /* mpiRankI is the MPI rank value: 0 to (N-1) */
    loc = 3 - (mpiRankI % 3); /* 0 reserved for OS & Hypervisor */
    attr = ALLOC_INIT;
    alloc_set_node_preferred( &attr, loc );
    /* msp must be global! */
    msp = create_mspace_with_attr( 0, 0, &attr ); 
}

\end{verbatim}
\end{minipage}

The standard 750$\,\times\,$750 pixel test image 
was used to measure the performance of CRBLASTER 
with 1 to 45 tiles
using all four of the memory controllers of the Maestro processor 
(see the filled squares in Fig.~3).
The speedup of CRBLASTER using the L.A.COSMIC algorithm with 
9, 24, and 45 tiles was 
6.6, 11.6, and 13.0, respectively.
The computational efficiency with 9, 24, and 45 tiles was, respectively,
73.0\%, 48.4\%, and  28.8\%.

Using all 4 memory controllers improved the computational efficiency
of CRBLASTER  when a moderate number of tiles was used.
The computational efficiency of CRBLASTER using the L.A.COSMIC algorithm with 
3 tiles jumped from 83.5\% to 92.6\% when each subimage  used its own
memory controller instead of all sharing the same memory controller.
Similarly, when 6 tiles were used, the computational efficiency jumped from
61.4\% to 85.5\% when each memory controller worked with no more than
two unique subimages.

The L.A.COSMIC algorithm is memory intensive and at a certain point when too many
tiles are sharing only 3 memory controllers there will be too many memory
access conflicts (cache hits) and computational efficiency will suffer.
Note that with the {\em{mspace}} version of CRBLASTER using the L.A.COSMIC
algorithm, the computational efficiency for 45 tiles was 28.8\% which was slightly
down from 30.9\% when only one memory controller was used in the standard version.

Off-chip memory access is expensive and too many conflicts negatively 
impact computational efficiency. Ideally, one would like to double the size of the
L2 cache from 64 Kbytes \citep{Suh+2011}
to 128 Kbytes on each of Maestro's 49 tiles. Unfortunately,
that would have almost double the die size of the Maestro processor which already was
the largest chip ever fabricated by Boeing SSED using 90-nm CMOS 9SF technology.

Figure 3 also gives the performance of the CRBLASTER computational framework with a 
computationally-intensive algorithm.  The 5-pointed stars in that figure 
show the speedup and computational efficiency
when Poisson deviates with a mean of 137 were generated for
every pixel in the 750$\,\times\,$750 standard input image.  
Details of the Poisson deviate
algorithm used are given in Mighell (\citeyear{Mighell2010}).  The CRBLASTER computational
framework with a computationally-intensive algorithm has computational efficiencies that are
very good with large numbers of tiles: 82.8\% and 78.0\% for 36 and 45 tiles, respectively.
Measured computational efficiencies with the CRBLASTER computational framework
can be even better when using applications that have better compute/communicate ratios
(e.g., setting the mean of the Poisson distribution to 1000 instead of 137).

The effect of adding an FPU to each tile can be measured by building CRBLASTER with the
Poisson deviate algorithm for both the TILE64 and the Maestro platforms and comparing those
assembled codes on the MDB.
All floating point operations on TILE64 code are emulated in software whereas the FPU 
attached to each tile on the Maestro chip provided hardware assist to floating point computations
with Maestro code.
Ideally this test is done using only one tile as that eliminates time lost due to tile-to-tile communication.
Experiments show that CRBLASTER using its Poisson deviate algorithm is $\sim$11 times faster 
using the FPUs than when all floating point operations are emulated in software.  This test provides
a good upper limit to the performance boost one can expect from using the Maestro FPUs as most
applications will not be as computationally intensive as this particular application.

\section{Overheating the Maestro processor}

During the process of collecting the results reported in the previous section, the 350-MHz MDB would 
occasionally crash for no apparent reason.
During long runs using scripts that would execute CRBLASTER 45 times using 1 to 45 tiles
simultaneously, the CRBLSTER would crash after a random-appearing number of trials. Sometimes
an illegal instruction was encountered.  Sometimes a double fault interrupt was reported.
Those frequently could be recovered from but typically only for a few more trials.  Then generally a 
much more serious fault\footnote{Kernel panic - not syncing: Aiee, killing interrupt handler!}
would occur and the MDB would crash which would require a physical reboot of the MDB and
a 20+ minute upload of the bootrom before debugging could proceed.  Frequently, once the
reboot had taken place, the MDB would react perfectly.

I was telecommuting from NOAO's headquarters in Tucson, Arizona to the MDB at the Arlington, 
Virginia site of USC/ISI.
In order for my experiments to not interfere with with work of the USC/ISI staff,
I would typically start using the MDB at 3:00 PM in Tucson which would be 6:00 PM
in Arlington.  After a particularly bad run of crashes, Mikyung Kang (USC/ISI) tried to recreate
my crashes but could not do so --- during the standard working day.  One night, she tried to
repeat the experiment at 9:00 PM and found that then
the MDB crashed on her too.

A little investigation revealed that the air conditioning to the room where the MDB was
located was turned off every day 
at 5:45 PM in order to save energy.  
Joseph Suh and Dong-In Kang (USC/ISI) 
moved the MDB to another room which had air conditioning all 
the time (24 hours / 7 days a week).  
Scripts running CRBLASTER from 1 to 45 tiles simultaneously then ran perfectly with no
crashes at all.

Suspecting that the Maestro processor was itself overheating, I tried a new experiments.
By using CRBLASTER to do computationally intensive work with TILE64 code (forcing
the emulation of all floating point operations), I could cause the MDB to fail at its new
location.  Moving the MDB to colder locations within the computing room caused the
crashes to go away --- until more intensive applications were run.
Eventually the MDB had to be moved to be directly in front of the room's
air conditioner unit in order to get the most stable performance out of the MDB.

It might be possible that heat was building up somewhere on the MDB external to the Maestro
processor itself.  However, it is most likely that heat was gradually building up in the central portion
of the chip and ---  if run intensively over a long period of time --- would eventually cause
the chip to overheat and crash.  Based on the reported errors, it appears that the chip got so hot
that random bits were being flipped within registers or inside instructions.  Neither of which
is good for the standard operation of any processor.  Apparently no permanent damage
to the Maestro processor occurred when the chip overheated.  The operating system would
crash the MDB, which allowed it to cool off sufficiently during the many minutes 
required for uploading a bootrom.

Assuming that the overheating problem is not unique to the MDB that I used,
overheating the Maestro processor can be mitigated several ways.
A mechanical solution would be to replace the current passive heat sink with a heat pipe.
One could reduce the amount of excess heat generated 
by operating the MDB at a slower clock speed.
Boeing SSED is now considering setting the clock speed for general MDBs to be between 133 -- 275 MHz.
Alternatively, the operating system could be made to be aware of the chip thermal
environment and have the chip halt operations whenever  the temperature of the central portion
of the Maestro chip exceeds a critical value.  

\section{Conclusion}

The speedup factor of CRBLASTER with the L.A.COSMIC algorithm was 13.9\break
using 45 tiles simultaneously
--- giving the equivalent peak performance of a  4.8 GHz processor when run on a 350-MHz 
Maestro Development Board with an Interim Test Chip.

The performance of CRBLASTER running a memory-intensive application can be improved significantly,
for a moderate number of tiles ($N < 20$), when user-defined memory heaps are used that
are associated with all four of the memory controllers on the Maestro processor.  

This project demonstrates the value of using real scientific applications during the testing phase of next-generation flight computer hardware; complex real-world scientific applications can stress hardware in novel ways 
that may not necessarily be revealed while executing simple applications or unit tests.

Although it is still early days for the Maestro processor and Maestro Development Boards, 
it is now clear
that the Maestro processor has the potential to be an enabling\break
space-flight computing technology for the next generation
of U.S. Government\break satellites and NASA astrophysical missions.

\bigskip
\acknowledgements 
I would like to thank the Program Organizing Committee 
for the invitation to make this presentation
and the Local Organizing Committee for the logistical support they provided.
I would also like to thank the following people for helping making this research possible:
Marti Bancroft, 
Steve Crago, 
Jinwoo Suh, 
Dong-In Kang,
and
Mikyung Kang.
This work was supported by two grants
from the
National Aeronautics and Space Administration,
NASA Grant Numbers
NNG06EC81I
and
NNX10AD45G,
from the 
Applied Information Systems Research (AISR) Program 
and the
Astrophysics Data Analysis Program (ADP),
respectively, 
of NASA's Science Mission Directorate.

\bibliography{I07}

\begin{thebibliography}{}
\expandafter\ifx\csname natexlab\endcsname\relax\def\natexlab#1{#1}\fi
\expandafter\ifx\csname url\endcsname\relax
  \def\url#1{\texttt{#1}}\fi
\expandafter\ifx\csname urlprefix\endcsname\relax\def\urlprefix{URL }\fi
\providecommand{\eprint}[2][]{\url{#2}}

\bibitem[{{Mighell}(2011)}]{Mighell2011}
{Mighell}, K. {2011}, in {Infotech@Aerospace 2011 (AIAA-2011-1448)}

\bibitem[{{Mighell}(2010)}]{Mighell2010}
{Mighell}, K.~J. 2010, \pasp, 122, 1236

\bibitem[{{Suh} et~al.(2011){Suh}, {Mighell}, {Dong-In}, \& {Crago}}]{Suh+2011}
{Suh}, J., {Mighell}, K.~J., {Dong-In}, K., \& {Crago}, S.~P. 2011, 2012 IEEE
  Aerospace Conference, Big Sky, Montana, March 2012 (submitted)

\bibitem[{{Tilera}(2010)}]{Tilera2010}
{Tilera} 2010, Tilera Multicore Development Environment Application Libraries
  Refference Manual, {Doc.~No.~UG227 (Release 2.1.0.98943, April 1, 2010),
  Tilera Corporation}

\bibitem[{{Tody}(1986)}]{Tody1986}
{Tody}, D. 1986, Proc.~SPIE, 627, 733

\bibitem[{{Tody}(1993)}]{Tody1993}
--- 1993, Astronomical Data Analysis Software and Systems II, 52, 173

\bibitem[{{van Dokkum}(2001)}]{vanDokkum}
{van Dokkum}, P.~G. 2001, \pasp, 113, 1420

\bibitem[{{Wells} et~al.(1981){Wells}, {Greisen}, \& {Harten}}]{Wells+1981}
{Wells}, D.~C., {Greisen}, E.~W., \& {Harten}, R.~H. 1981, \aaps, 44, 363

\end{thebibliography}

\end{document}